\documentclass[twocolumn]{svjour3}          
\usepackage{graphicx}
\usepackage{epsfig} 
\usepackage{amsmath}
\usepackage{amsfonts}
\usepackage{amssymb}
\usepackage{multirow}
\usepackage{indentfirst}
\begin{document}

\title{Neighboring Interactions in a Periodic Plasmonic Material for Solar-Thermal Energy Conversion}

\author{Terence D. Musho$^{\rm a}$ $^{\ast}$\thanks{$^\ast$Corresponding author. Email: tdmusho@mail.wvu.edu
\vspace{6pt}} and Anitesh A. Lal$^{\rm a}$  and  Zackary J. Coppens$^{\rm b}$\\\vspace{6pt}  $^{a}${\em{Department of Mechanical and Aerospace Engineering, West Virginia University, Morgantown, WV 26506-6106, USA}};
$^{b}${\em{Department of Mechanical Engineering, Vanderbilt University, Nashville, TN 37235, USA}}\\ }

\maketitle


\begin{abstract}
A periodic plasmonic meta-material was studied using finite-difference time domain (FDTD) method to investigate the influence of neighboring particles on the near unity optical absorptivity. The meta-material was constructed as a silver nanoparticle (20-90nm) situated above an alumina (Al$_2$O$_3$) dielectric environment. A full parametric sweep of the particle width and the dielectric thickness was conducted. Computational results identified several resonances between the metal-dielectric and metal-air that have potential to broadening the response through stacked geometry. A significant coupled resonance between the metal-dielectric resonance and a cavity resonance between particles was capture as a function of dielectric thickness. This coupled resonance was not evident below dielectric thicknesses of 40nm and above cavity widths of 20nm. Additionally, a noticeable propagating surface plasmon polariton resonance was predicted when the particle width was half the unit cell length. 
\end{abstract}

\section{Introduction}
Over the past decade there has been a tremendous interest in manipulating optical wavelengths by tailoring the typology of plasmonic meta-materials (MMs)~\cite{henzie07}. The tailoring of electromagnetic MM's have demonstrated augmented optical properties such as negative index of refraction, artificial permittivity, and permeability. More recently, research of meta-materials has focused on two extremes of optical wavelength manipulation. These two extremes include, 1) the development of meta-materials with low-loss characteristics for optical switching applications~\cite{boltasseva11,xiao10} and 2) the development of meta-materials with nearly perfect absorption properties for energy conversion applications~\cite{hedayati11,ju11}. The following research focuses on the latter by investigating the influence of geometrical aspects on perfect absorbers and understanding the neighboring interactions with future application for solar thermal energy conversion. 

One of the difficulties in understanding the response of a perfect optical absorbing meta-material is there are several geometry parameters that are critical to maintaining the desired response. Furthermore, as will be discussed in this study, as neighboring nanoparticles are brought close together, competing interactions act both destructively or constructively to either enhance or diminish the response. By understanding the neighboring influences on the optical response there is a potential to broaden the perfect absorption characteristics. In the proposed application of this study, this equates to increase solar to thermal conversion efficiency. 

To limit the complexity of this study the plasmonic material geometry will be limited to a simple metallic nanoparticle situated on top of a dielectric thin film followed by a thick ground plane. This geometry has been demonstrated both experimentally and theoretically~\cite{hao11,landy11,qui12} to have a nearly perfect absorption response. Many of these authors also demonstrated that by controlling the particle diameter and dielectric thickness the response of these very simple devices can be tuned to particular optical frequencies~\cite{tuong13,tuong12,wakatsuchi10}. By taking the concept of the single particle perfect absorber and introducing the neighboring particle it is hypothesized that a third degree of freedom will allow the response to be broadened. 

\begin{figure} [!ht]
\centering
\includegraphics[width=\columnwidth]{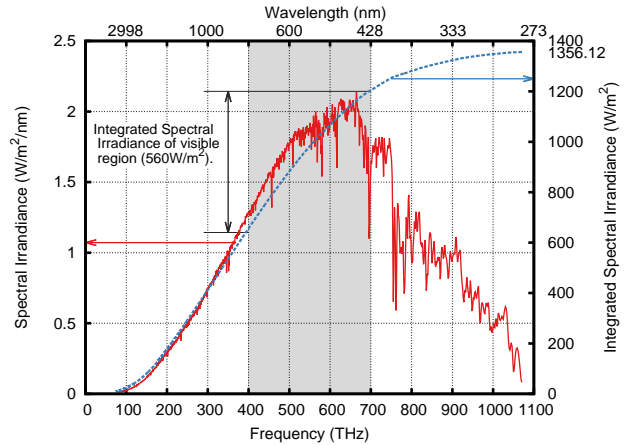}
\caption{Plot of the solar spectral irradiance as a function of frequency regenerated from data obtained from ASTM 490~\cite{ASTM490}. The right axis correspond to the integrated spectral irradiance, which has a cumulative total of 1356.12W/$m^2$.}
\label{fig:solarirradiance3}
\end{figure}

Meta-materials have been both experimentally and computationally shown to respond to all wavelengths of electromagnetic radiation~\cite{alici11} but the visible region has garnered a lot of interest for its applicability to many optical applications. Because our objective is to design a solar absorber the frequency of interest will be limited to the visible regime, that is, from 428-750nm (400-700 THz). Justification for focusing on this region initially can be reasoned based on a plot of the solar irradiance outside the earth's atmosphere as show in Figure~\ref{fig:solarirradiance3} and the corresponding left axis. The right axis of the same figure is the integrated spectral irradiance and can be more easily interpreted as the maximum heat flux if there is complete absorption of the incident electromagnetic radiation. The same figure also brings emphasis to the integrated irradiance of only the visible region, which is 560$W/m^2$. The visible region contributes to nearly 50\% of the total heat flux when compared to the total heat flux of 1356.12$W/m^2$ for the whole spectrum. 

In comparison to conventional materials such as carbon black and graphite, which can absorb heat in excess of 85\%~\cite{mizuno09} it is reasoned that a broadband meta-materials could potentially absorb heat in excess of 90\% as demonstrated at single frequencies~\cite{tung10,aydin11,hao11}. Furthermore, a limitation of carbon based materials stems from the reflectivity at air-carbon interface and the resulting emissivity of approximately 0.8. Recent approaches to use carbon nanotubes~\cite{mizuno09} have increased the emissivity upwards of 0.9 but there are practicable limitations to adhering nanotube to a substrates and providing good thermal transport across the interface. A unique advantage of the proposed metal-dielectric materials is the thermal barrier of the interface is avoided by exciting plasmons within the dielectric or alumina medium. Furthermore, alumina has an inherently large thermal conductivity (40$W/mK$ at 20C) due to the large phonon contribution to the thermal conductivity that will aid in the dissipation of localized dielectric heating.

In this study it is assumed that the incident electromagnetic energy will be absorbed in the form of surface oscillating plasmons. Followed by the assumption that the plasmons will decay into lattice vibration or phonons and diffuse through the substrate of the device. Often this decay of plasmons to phonons is referred to as dielectric heating but can also be related to Joule heating. By assuming the material is dispersive and that the magnetic response is negligible, it is possible to write a simple expression that relates the complex permittivity of a material to the volumetric energy generation. In order to utilize this expression, it is assumed that the material undergoes a forced harmonic oscillation with an incident electric field. Therefore, a material's electronic response to a polarizing field can be characterized with the following, $D=\epsilon_0\vec{E}+\vec{P}=\epsilon_0\epsilon\vec{E}$. This expression governs that the displacement ($D$) of the electrons is proportional to the applied electric field ($E$) and the polarization ($P$). The expression can be utilized to quantify the volumetric heating through Poynting's theorem, which is derived in the literature~\cite{hao11}. Under the harmonic assumption, the time averaged expression for the volumetric heat generation within the material will take the following form,

\begin{center}
\begin{equation} 
\dot{q}=\frac{1}{2} \epsilon_0 \omega \text{Im}(\epsilon(\omega)) \|\vec{E}\|^2,
\end{equation}
\end{center}

where $\epsilon_0$ is the permittivity of free space, $\omega$ is the angular frequency, $ \|\vec{E}\|$ is the norm of the electric field vector, and $\text{Im}(\epsilon(\omega))$ is the imaginary part of the permittivity of the dispersive material. The units of this expression are in power (work done per unit time) per unit volume and the general form of the equation can be related to Joule heating ($P(t)=I(t)V(t)$). This previous expression can be integrated over volume to determine the total power absorbed by the material at a particular frequency.

\begin{center}
 \begin{equation}
q= \int_V  \frac{1}{2} \epsilon_0 \omega Im(\epsilon(\omega)) \|\vec{E}\|^2 dV
\end{equation}
\end{center}

\section{Plasmonic Material Design}
\label{sec:method}
\begin{figure}[!ht]
\centering
\includegraphics[width=0.94\columnwidth]{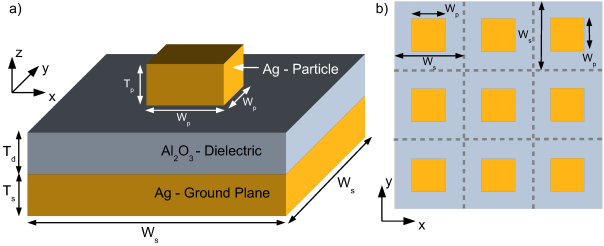}
\caption{Dimensional illustration of the meta-material nano-antenna constructed from a nanoparticle separated from a ground plane with an alumina dielectric. Critical parameters to this research are the particle width ($W_p$) and dielectric thickness ($T_D$). The meta-material composite is perfectly symmetrical and infinite in the xy-plane.}
\label{fig:dimension3d(2)}
\end{figure}

To limit the complexity of the antenna, a very simple metal-dielectric-metal construction was implemented as illustrated in Figures~\ref{fig:dimension3d(2)}.  A unit cell representation was used to alleviate the computational expense of the simulation while still maintaining an accurate response of the whole material through the use of Bloch boundary conditions at the minimum and maximum x- and y-boundaries. The geometry of the nanoparticle anchored on top of a dielectric thin film was square with extent in x- and y-dir being equivalent to maintain the square geometry throughout the study. The thickness of the particle was held at 50nm in the z-dir.  Below the dielectric material a thick ground plane (substrate) is situated with the same composition of the top nanoparticle. The remaining space between the top of the material and the upper port was assumed vacuum region. 

The top nanoparticle and the bottom ground plane for the following simulations were composed of silver ($Ag$). The dielectric thin film situated between the top particle and the bottom ground place was chosen to have a composition of alumina or aluminum oxide (Al$_2$O$_3$). The substrate had dimensions $W_G \times L_G \times T_G$ upon which the alumina with identical lateral dimensions $W_G \times W_G \times T_D$ was constructed. The silver nano-antenna with dimensions $W_p \times L_p \times T_p$ was located in the geometric center on top of the alumina layer. By controlling the extent of the substrate lateral dimension $W_G \times L_G$ the distance between neighboring particles was controlled. 

The boundaries that lie in the y-z and x-z plane were periodic such that the transverse magnetic field in the y-z plane was zero and the transverse electric field in the x-z plane was zero.  A polarized TEM plane-wave was propagated in the  $-z$ direction from the open boundary in the x-y plane in the maximum z location of the computational domain. The bottom of the computational domain in the minimum z location had an open boundary condition. The scattering parameters were monitored at both of the boundaries in the z-dir.

A Drude model was employed because the silver material is dispersive. Table~\ref{tab:drude} provides a list of Drude parameters for silver. The alumina was assumed loss-less with a fixed permittivity of 6.20, which is equivalent to a refractive index of 2.49 ($n=\sqrt(\epsilon)$). This value is based on experimental studies of alpha-phase alumina ($\alpha-Al_2O_3$)~\cite{french98}. The permeability of all the materials was assumed to be unity provided this study is focused on the optical region of the electromagnetic spectrum.

Because of the variation of the reported Drude parameters in recent literature, the Drude parameters were compared to experimental data and a least squares fit was used to determine the Drude parameters for this study. In order to plot the Drude parameters the permittivity was written in the real and imaginary form. By assuming the material response to incident radiation is harmonic the Drude model that describes the complex permittivity takes the following form,

\begin{equation}
\epsilon(\omega) = \epsilon_{\infty}- \frac{\omega_{p}^2}{\omega^2+i\omega\gamma_c} = \epsilon_1+i\epsilon_2,
\end{equation}
where $\omega_p$ is the plasma frequency, $\gamma_c$ is the collision frequency. The complex permittivity can be separated into a real part ($\epsilon_1$) and an imaginary part ($\epsilon_2$). Following the derivation of Ordal~\cite{ordal85}, the real part and imaginary part take the following form,

\begin{equation}
\epsilon_1 = \epsilon_{\infty}- \frac{\omega_{p}^2}{\omega^2+(2\pi\gamma_c)^2},\\
\end{equation}
\begin{equation}
  \epsilon_2 = \frac{(2\pi\gamma_c)\omega_{p}^2}{\omega(\omega^2+(2\pi\gamma_c)^2)}.\\
\end{equation}

\begin{figure}[!ht]
\centering
\includegraphics[width=\columnwidth]{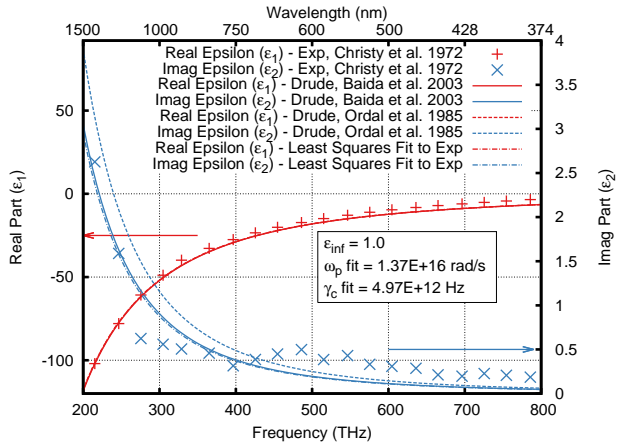}
\caption{Plot of the real and imaginary permittivity comparing the Drude model to experimental data~\cite{christy72}. The inset parameter ($\omega_{p}$fit and $\gamma_{c}$fit) were determined through a least squares fit of Christy's experimental data.}
\label{fig:drude}
\end{figure}

Using the equations for $\epsilon_1$ and $\epsilon_2$ the data provided in the experimental research of Christy~\cite{christy72} was fit with free parameter $\omega_p$ and $\gamma_c$. The result of this fit along with common parameters cited from other studies are depicted in Figure~\ref{fig:drude} and the corresponding Table~\ref{tab:drude}. It is mentioned by Christy, there is a large error in $\epsilon_2$ in the infrared region when the wave vector ($k$) is large compared to n, stemming from the relation $\epsilon_2=2nk$. Christy goes on to mention that the error in $\epsilon_2$ is less significant in the optical region where n and k are proportional. Therefore, for this study the experimental data is used with mild confidence but the reader should be aware of the apparent difference between $\epsilon_2$ fit using the Drude model and the experimental data.

\begin{table*} [!ht]
\begin{center}
  \begin{tabular}{ l | c | c }
  \hline
\hline
    Source & Plasma Freq. ($\omega_p$) [rad/s] & Collision Freq. ($\gamma_c$) [Hz] \\ \hline
    \hline
    Baida et al.~\cite{baida03}  & 1.37x10$^{16}$ & 5.10x10$^{12}$  \\ \hline
    Ordal et al.~\cite{ordal85}  & 1.37x10$^{16}$  & 6.57x10$^{12}$ \\ \hline
    Fit to Christy et al.~\cite{christy72}  & 1.37x10$^{16}$  & 4.97x10$^{12}$\\ \hline
  \end{tabular}
  \caption{Dispersion parameters for a Drude dispersion model to describe the dispersive nature of the silver material.  All materials were assumed to have a permeability of unity.}
  \label{tab:drude}
\end{center}
\end{table*}

\section{Computational Model}
\label{sec:computational}

A finite-difference time-domain (FDTD) computational method employed in the CST Microwave Studio Package was used to predict the scattering parameters as a function of neighboring interactions and geometry of the antenna. To aid in relating the scattering parameters to physically measurable quantities the silver substrate of the device was constructed sufficiently thick to eliminate the transmission of electromagnetic energy.

The absorptivity and the reflectivity from the view point of the top most boundary (port) in the positive z-dir above the device were predicted through the scattering parameters at that port. Because the silver substrate had a sufficient thickness the scattering parameter from the top port to the bottom port (S$_{12}$) was assumed and confirmed by the simulation to be zero. This resulted in simplification of the scattering parameters' relation, $S_{11}+S_{12}=1$. Defining an energy balance where, A($\omega)$ + R($\omega$) + T($\omega$) = 1, where A($\omega$) is the absorptivity, R($\omega$) is the reflectivity and T($\omega$) is the transmissivity, and assuming the transmissivity is zero the energy balance simplified to the following,  A($\omega)$ + R($\omega$) = 1. The reflectivity then can be defined as scattering parameter $R(\omega)=\left|S_{11}\right|^2$ and hence the absorptivity can be defined as A($\omega)$ =1-$\left|S_{11}\right|^2$.

A tetrahedral mesh was used to discretize the computational domain. The FDTD solution was specified to have ten grid points per wavelength. Over the frequency range up to 100 points were simulated and based on those points a 1001 point spline was fit to provide a smooth response over the range of interest. Also, the adaptive mesh refinement feature was used in CST to locally refine mesh cells and check grid mesh independence from the solution. 

\section{Discussion}
\label{sec:discussion}
This research was divided into two areas of interest, 1) study of the optimal dielectric thickness and particle diameter for a fixed domain of 91nm and 2) study of the influence of neighboring particles with varying domain length. The following section will begin with the optimization study of particle width and dielectric thickness.

\subsection{Absorptivity as a Function of Particle Width and Dielectric Thickness}
\label{sec:optimization}
\begin{figure}[!ht]
\includegraphics[width=1.1\columnwidth]{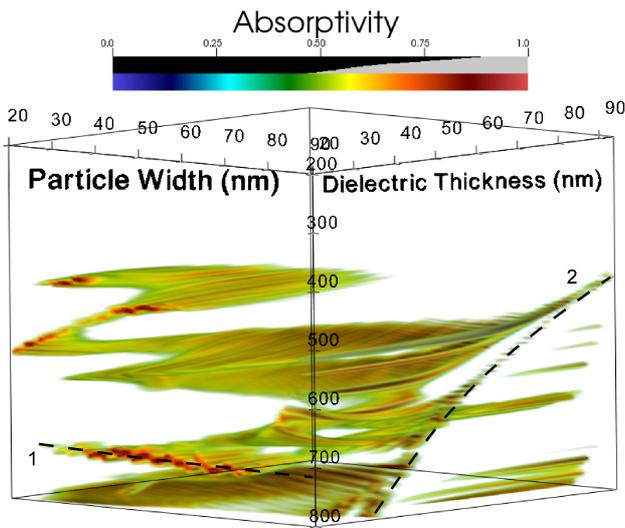}
\caption{Plot of the absorptivity as a function of particle width (W$_p$) and the dielectric thickness ($T_d$) at a fixed domain length of 91nm ($W_s$). Only points with an absorptivity above 0.5 are shown in the figure. Dashed line 1 is the response due to the metal-air. Dashed line 2 is a coupled response of the metal-dielectric and the neighboring cavity resonance. At a dielectric thickness of 90nm the dashed line approaches the LSSP peak of the metal-dielectric. Cross-sections from this 3D plot can be found in Figure~\ref{fig:ptsyz} at a constant dielectric thickness and Figure~\ref{fig:ptsxy} at a constant particle width.}
\label{fig:optim}
\end{figure}

The absorptivity as a function of particle geometry and dielectric thickness was investigated by conducting a full parametric design sweep of the space. The approach taken was to use the FDTD model to evaluate a series of 10,000 design points with the space to determine the trends of the design space to guide in the design of meta-material with peak absorptivity. This parametric study was carried out by fixing the domain width to 91nm. The first degree of freedom in the parametric sweep was the particle width ($W_p$), which was varied from 20nm to 90nm  for 100 samples. The second degree of freedom was the dielectric thickness ($T_d$), which was varied from 20nm to 100nm for 100 samples. The absorptivity was calculated for each configuration point and the results are shown in Figure~\ref{fig:optim}, which illustrate the response of the peak absorption as a function of the two degrees of freedom.

Figure~\ref{fig:optim} provide a complete view of the particle and dielectric response for the simple antenna. Captured in this figure is the response of not only the localized surface plasmon polaritons (LSPPs) due to both the metal-dielectric response and the metal-air response but also propagating SPPs (PSPPs). There are two trends that are most obvious and they lie on either plane of particle width v. frequency and dielectric thickness v. frequency. The first trend is outlined by the dashed line and the number 1 in the particle width v. frequency plane. This response is independent of the particle width and investigating the spatial electric field plot determined the FDTD simulation the field is localized at the top corners of the particles suggesting this a response due to the metal-air interface, where we assumed air has an electric permittivity of unity. The second trend on the adjoining plane of dielectric thickness v. frequency is a more interesting trend as it pertains to the metal-dielectric response.

\begin{figure}[!ht]
\centering
\includegraphics[width=0.94\columnwidth]{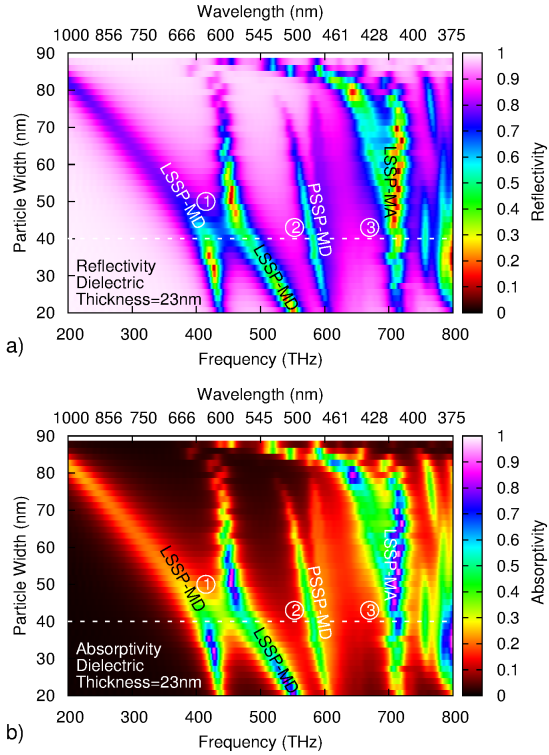}
\caption{Plot of the (a) reflectivity and (b) absorptivity as a function of particle width and frequency. The dielectric thickness is T$_d$=23nm. Number (1) indicates localized surface plasmon polaritons between the metal and dielectric (LSSP-MD). Number (2) is a resonance of propagating surface plasmon polaritons at the metal-dielectric interface (PSSP-MD). This PSSP-MD is peak when the particle width is half the domain length. Number (3) is a resonance of localized SSP at the metal-air interface (LSSP-MA)}
\label{fig:ptsxy}
\end{figure}

\begin{figure}[!ht]
\centering
\includegraphics[width=0.64\columnwidth]{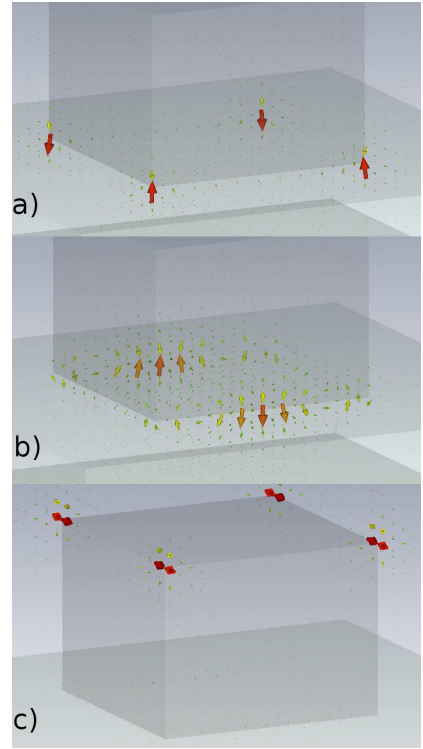}
\caption{Plot of the electric field at three points of frequency that correspond to the three areas of emphasis in Figure~\ref{fig:ptsxy}. (a) is the LSSP-MD at 416THz, (b) is PSSP-MD at 582THz and, (c) is LSSP-MA at 714THz.}
\label{fig:ptsxy_efield}
\end{figure}

To aid in the interpretation of Figure~\ref{fig:optim} two different 2D slices have been taken. The first slice is in a dielectric thickness v. frequency plane at a fixed dielectric thickness of 23nm, see Figure~\ref{fig:ptsxy}. Within this slice there are three distinct region as labeled in the figure. The first region looking from left to right in the figure is the LSPPs that are at the four corners of the particle at the interface of the metal and dielectric (LSSP-MD). Following the 40nm particle width line, emphasized as a white dashed line in the figure the left most branch (416THz) is associated with a counter-clockwise circular current ninety degrees out of phase with the incident signal. The neighboring LSSP branch (497THz) is a clockwise circular current ninety degrees out of phase with the incident signal. In both of these cases the heating is subsequently localized at the corners of the nanoparticle within the dielectric. Following further right to 582THz there is a propagating SPP resonance that propagates ninety degrees out of phase along the y-dir. The heating at this resonance would be along the x-min and x-max edge of the nanoparticle at the metal dielectric interface. This is an advantageous resonance from a thermal point of view as there would be an induced thermal gradient within the dielectric. The final area of interest along the line at (3) is the resonance due to the metal-air. All three of these resonances are visualized by electric field vectors in three dimensions in Figure~\ref{fig:ptsxy_efield}.

Focusing on the first point of interest in Figure~\ref{fig:ptsxy}, the dependence of the LSSP on the particle width provides evidence for a stacked design that would result in broadband absorptions. If you were to form a pyramidal stack~\cite{cui12} of alternating metal-dielectric layers where the dielectric was maintained a 23nm and the metal width was varied from 20nm to 60nm, the frequency range between 400THz and 550 THz could be absorbed. More interesting, there is a noted blue shift in the PSSP-MD resonance (Number 2 in Figure~\ref{fig:ptsxy}) as a function of dielectric thickness, which suggests that several 40nm particles atop increasing thickness of dielectric (20-90nm) would encompass a frequency range between 575THz and 416THz. Furthermore, inducing a thermal gradient across the plasmonic device.

\begin{figure}[!ht]
\centering
\includegraphics[width=0.94\columnwidth]{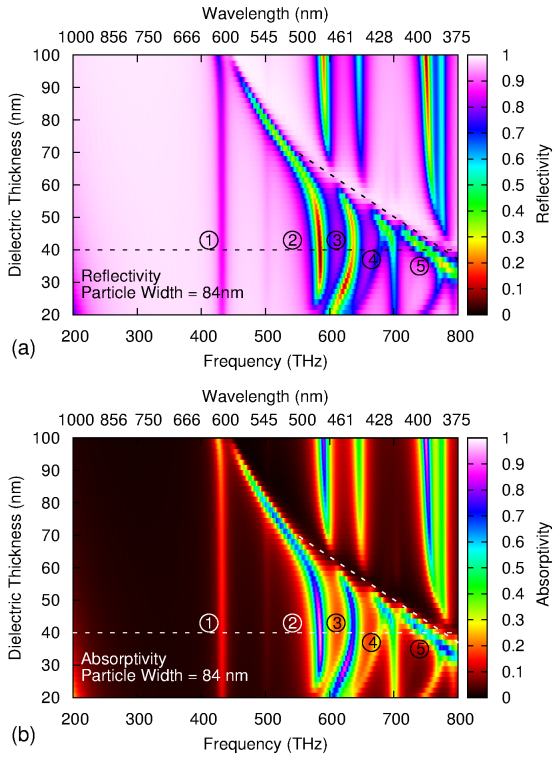}
\caption{Plot of the (a) reflectivity and (b) absorptivity as a function of dielectric thickness and frequency. The particle width is W$_p$=84nm. Number (1) is a weak resonance of the fundamental LSSP between the metal and dielectric. Number (2) is a cavity resonance, (3) is a PSSP resonance, (4) is a LSSP between the metal and air, and (5) is a second order cavity resonance. The  angled dashed line above a dielectric thickness of 40nm is a result of coupling between the cavity resonance and metal-dielectric resonance.}
\label{fig:ptsyz}
\end{figure}

The second slice of interest from the 3D plot in Figure~\ref{fig:optim} is in a dielectric thickness v. frequency slice at a particle width of 84nm, see Figure~\ref{fig:ptsyz}. Here we gather another range of resonances similar to what was shown in the previous slices. However, a strong dependence on dielectric thickness is noted and emphasized by the dashed line. Again, if we concentrate our discussion to a dielectric thickness of 40nm and start from the left and increase in frequency the first peak, which is not very significant at this geometry, is the LSSP resonance at 416THz. This LSSP resonance is not significant in intensity and lacks a dependence on the dielectric thickness. If a vector plot of electric field were investigated there would be a localized field at the corners of the nanoparticle. Moving right, at (2) the resonance (581THz) is an interaction between the neighboring particles within the cavity, which is often referred to as a cavity resonance. (3) is a PSSPs resonance that is influenced by the dielectric thickness. (4) is a LSSP-MA at the metal-air interface and (5) is a second order cavity resonance with the neighboring particles. If the dielectric thickness is greater than 40nm there is a strong interference that leads to the fundamental LSSP resonance at 416THz. In reference to Figure~\ref{fig:optim} this interference is a result of the response of the nanoparticle and the ground plane and the nanoparticle and neighboring nanoparticles. As the particle width decreases the interference is not apparent and the fundamental LSSP resonance is only present.  

\subsection{Influence of Neighboring Particles on the Electromagnetic Response}
\label{sec:influence}
One of the complications with designing a periodic meta-material is the influence of neighboring structures. As noted in the previous section, the response of the material can be dependent on distance between neighboring particles. For this sub-study the influence of neighboring particles is discussed in terms of several known theories. In doing so, a series of simulations were conducted by increasing the distance between the edge of the particle and the periodic boundary condition for a single unit cell. The domain was fixed at 500nm and carried out in an identical manner as discussed in Section~\ref{sec:method}.

\begin{figure}[!ht]
\includegraphics[width=\columnwidth]{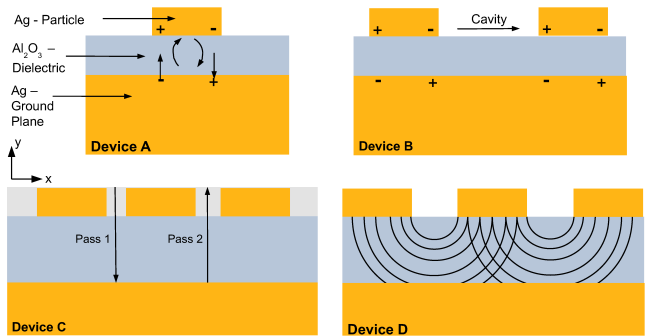}
\caption{Illustration of the four dominate interactions that are responsible for the response. Design A is an anti-parallel current design. Design B is a cavity interaction between particles. Design C is an effective medium design. Design D is a grating coupling design.}
\label{fig:design}
\end{figure}

\begin{figure}[!ht]
\includegraphics[width=\columnwidth]{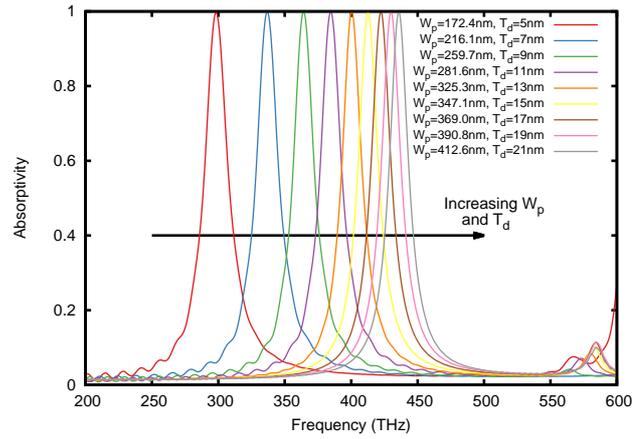}
\caption{Results from Design A of an anti-parallel perfect absorber. As the distance between particles increases the response red shifts.}
\label{fig:set1}
\end{figure}

A series of four designs illustrated in Figure~\ref{fig:design} were investigated. The first design involves varying both the particle spacing and dielectric thickness and defines the fundamental plasmon frequency. The geometries where the particle spacing and dielectric thickness have nearly perfect absorption is explained by anti-parallel current~\cite{Watts12}. Figure~\ref{fig:set1} shows the absorption parameters for a select number of cases. As the period size decreases the fundamental plasmonic frequency red shifts. This will serve as a baseline to identify the fundamental mode from the coming designs.

\begin{figure}[!ht]
\includegraphics[width=\columnwidth]{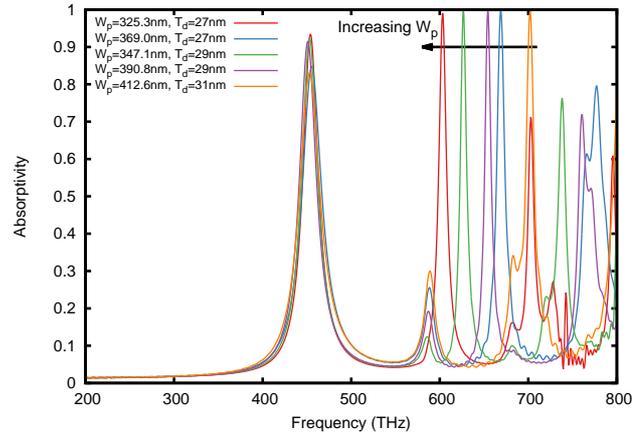}
\caption{Results from Design B of a cavity response between particles resulting in perfect absorption. As the cavity increases the response blue shifts.}
\label{fig:set2}
\end{figure}

The second design involves forming a cavity between the neighboring particles. This corresponds to Design B in Figure~\ref{fig:design}. The cavity response of this design is analogous to that of a nanohole array where the incident electromagnetic wave resonates between the particles. Because of the proximity of the closely spaced silver nano-particles, extreme coupling occurs. This coupling leads to strong field enhancement in the near field, thus enhancing the light-matter interaction and resulting in perfect absorption with multiple resonances.  Figure~\ref{fig:set2} shows select runs where the distance between the particles is decreased. As the distance decreases, the response red shifts. However, the fundamental mode remains stationary and the secondary mode is the peak that is shifted. This can be conceptually thought of as longer wavelengths coupling within larger hole diameters. Additionally, the dielectric thickness does not influence the response as illustrated by the selected points.

The third design is illustrated in Figure~\ref{fig:design} (Design 3). It is based on an effective medium theory. This theory presumes that the particles situated on top of the dielectric act as a singular material with material properties that depend on the volume fraction of silver and air. As the distance between the particles increases (particle size decreases) the volume fraction of air increases. Perfect absorption can be obtained by impedance matching the effective medium with the dielectric. A resonance of 768.5 THz occurs when the period is 85nm (a 10nm distance between particles) and the dielectric thickness is greater than 19nm. 

\begin{figure}[!ht]
\includegraphics[width=\columnwidth]{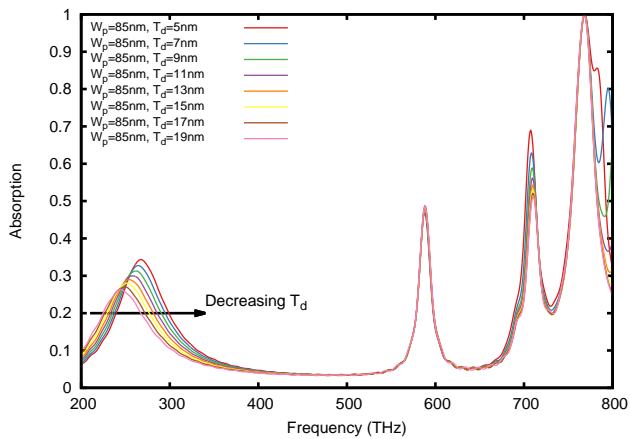}
\caption{Results of Design C of an effective medium material where the volume fraction of the silver and air impedance match the surrounding air.}
\label{fig:set3}
\end{figure}

To impedance match the effective medium with the dielectric, the ground plane was removed and a set particle geometry was simulated. The peak frequency with a reflection of zero marked the impedance matching at 768.5THz characterized by $Z=\sqrt{\frac{\mu }{\epsilon}}$, where $\mu$ equals unity.  Once this peak impedance-match frequency was known, the ground plane was put back in the simulation. The width of the domain was restricted to 85nm and the dielectric thickness was varied until the equivalent peak was found. The arrangement that demonstrated this similar response with the ground plane was a particle width of 85nm. This configuration resulted in the correct volume fraction of air and silver for the corresponding permittivity of air. A series of runs illustrate this in Figure~\ref{fig:set3}. The response of this effective medium requires that the dielectric thickness be greater than 19nm to facilitate a long enough traverse time to absorb the energy. 

\begin{figure}[!ht]
\includegraphics[width=\columnwidth]{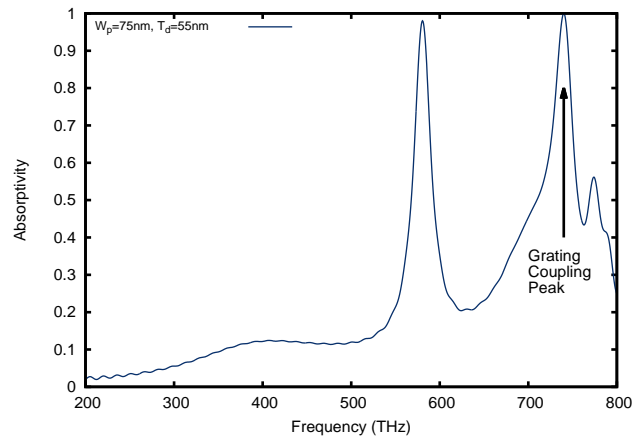}
\caption{Results of Design D of a grating coupler where the peak resonance is apparent when the spacing between the particles is twice the particle diameter.}
\label{fig:set4}
\end{figure}

The fourth design is a grating coupling design that is illustrated by Design 4 in Figure~\ref{fig:design}. The grating coupling is evident by removing the ground plane and still maintaining the optical response. The grating coupling is realized when the period of between the particles is twice the particle width. This is illustrated in Figure~\ref{fig:set4} where the particle width is set to 75nm and the particle period is 150nm. The result of the ground plane with the grating results in a perfect absorber.

\section{Conclusion}
By understanding the interaction between neighboring particles there is potential to design more efficient plasmonic devices but also the potential to broaden their response. In conducting a full parametric sweep of particle diameter and dielectric thickness at a fixed domain length of 91nm there were several resonances that could be attributed to the metal-dielectric and metal-air response.  At particle diameters within 20nm of the unit cell length, a coupled interaction between the cavity resonance and metal-dielectric resonance was noted. Similarly, for sufficiently thin dielectrics there was a dependence of the localized surface plasmon polaritons on the particle width at the corners of the nanoparticle and a strong resonance between the metal and air. The interactions of neighboring particles were reasoned to be explained by four current theories, which include 1) anti-parallel currents, 2) cavity resonances, 3) effective medium, and 4) grating coupling. Given the identification of these four interactions the influence on their geometry identified how the peak resonances shift.

\bibliographystyle{spphys}
\bibliography{journal}
 
\end{document}